# Measuring the Dynamical State of the Internet: Large Scale Network Tomography via the ETOMIC Infrastructure

Gábor Simon[a,b] József Stéger[a,b] Péter Hága[a]
István Csabai[a,b] Gábor Vattay[a,b]

[a] Department of Physics of Complex Systems, Eötvös Loránd University, Budapest, Hungary;
[b] Collegium Budapest Institute for Advanced Study, Budapest, Hungary

**Short Title**
Measuring the Dynamical State of the Internet

**Abstract**
In this paper we show how to go beyond the study of the topological properties of the Internet, by measuring its dynamical state using special active probing techniques and the methods of network tomography. We demonstrate this approach by measuring the key state parameters of Internet paths, the characteristics of queueing delay, in a part of the European Internet. In the paper we describe in detail the ETOMIC measurement platform that was used to conduct the experiments, and the applied method of queueing delay tomography. The main results of the paper are maps showing various spatial structure in the characteristics of queueing delay corresponding to the resolved part of the European Internet. These maps reveal that the average queueing delay of network segments spans more than two orders of magnitude, and that the distribution of this quantity is very well fitted by the log-normal distribution.

**Corresponding Author**
Dr. Gábor Simon
Department of Physics of Complex Systems,
Eötvös Loránd University
H-1518 Budapest Pf.: 32, Hungary.
Tel:+36 1 3722896, Fax: +36 1 3722866
E-mail gaba@complex.elte.hu

## Introduction

Internet is a complex network of computers and routers connected by direct wired or wireless links, where information is transmitted in discrete sized packets. Since 1970 the Internet evolved rapidly and in a decentralized manner from a few interconnected local area networks into a highly heterogenous global network spanning continents. The Internet provides various services and applications that became part of our every-day activity including web browsing, file transfer, multimedia, Internet-telephony, and others. In the future it is expected that the reliance of the population on the firm functioning of the Internet and on the new emerging applications will increase dramatically, thus it is of utmost importance to study and to understand the properties of this huge network at various levels of abstraction. By exploring the characteristics of the present-day Internet, and the various phenomena associated to its traffic it may be possible to extrapolate that knowledge to predict the properties of the future Internet and to foresee its problems.

Traditionally the Internet was studied by various engineering approaches that concentrated mostly on practical aspects, however recently with the accumulation of large datasets, it was recognized that the Internet can also be studied from a complex systems perspective, which represents a different level of abstraction. Complex systems [1] are large networks of interacting agents, where the emphasis of the studies are on the global properties, characterizing the system as a whole, by neglecting the fine details of the interactions. It is believed that microscopically vastly different complex systems have some universal features that are independent of the details. For example collective phenomena of congestion buildup are present both in the Internet and highway-traffic, showing analogous dynamics on a global scale [2]. Also both of these systems are characterized by self-similarity [3,4], and 1/f noise in the congested state [5-7].

Most of the existing work so far addressed questions concerning the topological properties of the Internet [8,9] or related topics like tolerance to attacks or failures [10]. These studies established interesting characteristics like the power-law degree distributions of the Internet connectivity graph on the autonomous system and router level. Although as the network evolves this continues to be an open issue and a hot topic, nevertheless it is essential to extend the study of the Internet beyond the discovery of topological properties towards the characterization of its dynamical state. Steps have already been taken in this direction; for instance in references [11,12] who revealed universal scaling in traffic intensity fluctuations.

From the point of view of the current, most widely used data transfer protocols like the variants of TCP, the most relevant state variables of the network are the loss-rates and the delays encountered by data packets on a path, since these quantities control the transmission rate of the transfer protocol. There are two different techniques to measure loss-rate and delay over the Internet, these are active probing and passive monitoring. Active probing involves injecting of probe packets into the network and analyzing the properties of the received probe-stream. This technique is flexible, has wide range of applicability, however its extensive usage may impose unnecessary load on the network. On the contrary passive monitoring does not impose any load, but it relies on the fact that the user must have an access to the network element under investigation, which is generally not the case. In the past years several measurement platforms have been developed for conducting active measurements over the Internet (e.g. Surveyor, Felix, AMP [13]), however these platforms can provide only end-to-end information between the participating nodes, where the measured characteristic can not be resolved on the parts of the end-to-end path. Also most of the existing active probing tools rely on extra cooperation of the routers in the path to process their packets. As the Internet continues to evolve towards more decentralized and heterogenous administration, in the future the cooperation of the network elements can be foreseen to be limited to the basic process of just storing and forwarding incoming probe packets.

The solution to these problems is provided by network tomography, which is a special class of active-probing measuring techniques, that is able to resolve the end-to-end delay statistics [14,15] and packet loss rates [16,17] to internal segments of the paths. In general a tomography measurement made from a single source to a set of receivers admits the determination of the delay statistics and loss-rates on each segment of an underlying logical tree that is spanned by the source and receiver nodes, and the branching nodes (nodes where the path of probes destined for different receivers diverge). By increasing the number of sources and receivers involved in a tomography measurement, the portion of the network for which state information can be resolved grows dramatically. Initially network tomography techniques were developed for the use with multicast probes [18-21], which requires the extra cooperation of the routers to support multicast functionality, however later these approaches were also extended to the case of using unicast probes [16,14,17], and performing the measurements from multiple sources [22,23], which makes unicast network tomography the most general tool to measure spatially resolved characteristics of an uncooperative Internet.

The main idea of unicast network tomography is to use back-to-back packet pairs, where each packet of a pair is destined to different receivers. As the packets of a pair traverse their paths, they experience the same network conditions on the common segment from the source to the branching node, which brings correlation into the time-series of the end-to-end characteristics. This inherent correlation property of such probe streams is the key to resolve the internal characteristics from the end-to-end measurements.

The delay experienced by a packet over an Internet path sums up from two non-negative components, a constant propagation delay and a time-varying component due to queueing in the buffers of routers. In this paper we are concerned with large-scale inference of queueing delay distributions in the Internet by performing extensive unicast network tomography measurements. The large-scale study of queueing delay distributions is motivated by the fact that this observable carries vast amount of information about traffic properties and the state of congestion on the measured path. By resolving the queueing delay distributions from end-to-end measurements we can draw a map of congestion of the network segments, analyze spatial structure, and identify highly congested or faulty segments.

The rest of the paper is organized as follows. In the following section we describe ETOMIC, the measurement platform, where the experiments were conducted. Next we proceed with the presentation of the results of our preliminary large-scale tomography measurement, while the detailed description of the queueing delay tomography method is given in the last section.

**The ETOMIC Measurement Platform**

To perform unicast queueing delay tomography in a real network environment poses several challenges. First in order to be able to measure true end-to-end delay, source and receiver nodes need to be synchronized to a common clock-reference, and must stay in the synchronized state during the measurements. Second, the measuring infrastructure has to be very precise in order to be able to resolve the microsecond-scale queueing delay components associated to high-bandwidth (multi-Gigabit) links. The precision of commercial workstations are insufficient for this task, thus to achieve sub-microsecond precision, a hardware solution is inevitable.

In the subproject of the European Union sponsored EVERGROW[24] Integrated Project, we are developing a state of the art high-precision, synchronized measurement platform, the Evergrow Traffic Observatory Measurement InfrastruCture (ETOMIC) [25]. This platform among others provides the ability to perform large-scale delay and loss tomography based on unicast probing techniques, and will be generally available and open to the public. Currently ETOMIC consists of 15 measuring nodes deployed at different locations in various European countries (See Table I., while in the future we plan to extend this number to 50 participating nodes. The measurement nodes and the network experiments with them are managed through a central management system that is accessible to the researchers through a web-based graphical user interface [26].

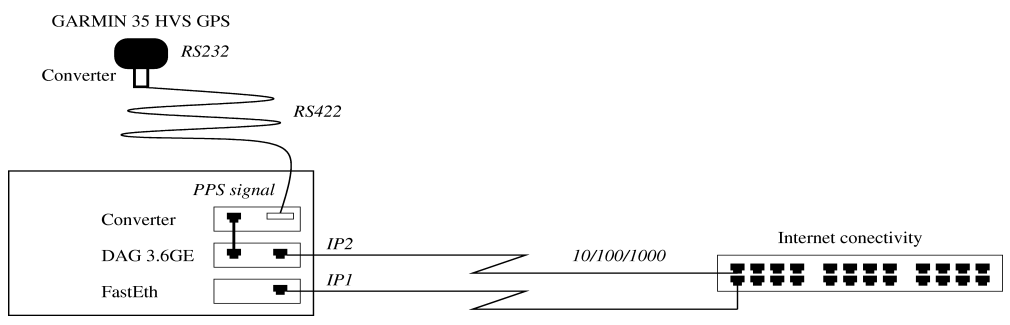

**Fig. 1.** The measuring node is a standard PC equipped with a DAG 3.6GE network interface card, connected to a GPS unit.

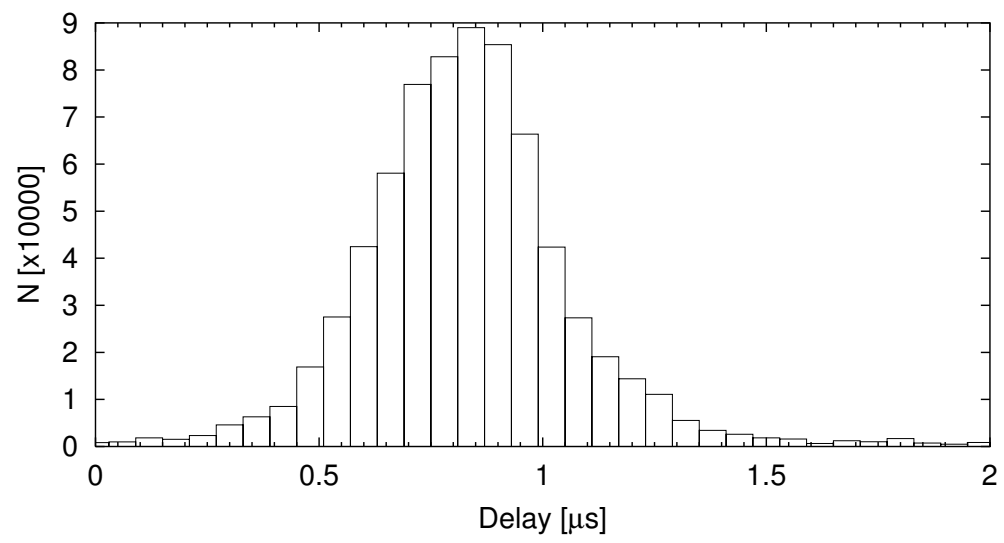

**Fig. 2**. One-way delay measurement between two gps-synchronized ETOMIC nodes over an empty link. The FWHM of the histogram indicates that the attainable precision of end-to-end delay measurements within ETOMIC is around 0.5µs.

**Table 1.** The list of current ETOMIC nodes

| ABBREVIATION | IP ADDRESS | LOCATION |
|---|---|---|
| SICS | 193.10.64.81 | Stockholm, Sweden |
| TELI | 217.209.228.122 | Stockholm, Sweden |
| ERIC | 192.71.20.150 | Stockholm, Sweden |
| UNAV | 130.206.163.165 | Pamplona, Spain |
| ASTN | 134.151.158.18 | Birmingham, England |
| HUJI | 132.65.240.105 | Jerusalem, Israel |
| OVGU | 141.44.40.50 | Magdeburg, Germany |
| ROME | 141.108.20.7 | Rome, Italy |
| UNIV | 193.6.205.10 | Budapest, Hungary |
| COLB | 193.6.20.240 | Budapest, Hungary |
| ELTE | 157.181.172.74 | Budapest, Hungary |
| UPAR | 193.55.15.203 | Paris, France |
| SALZ | 212.183.10.184 | Salzburg, Austria |
| UBRU | 193.190.247.240 | Brussels, Belgium |
| CRET | 147.27.14.7 | Chania, Greece |

The schematics of an ETOMIC measurement node is displayed in fig. 1. It is based on standard PC hardware, but also includes an Endace DAG 3.6GE card as the network monitoring interface, which is specifically designed for precise active and passive measurements [27]. These DAG cards provide very accurate time-stamping of the probe packets, with a time-resolution of 60 ns, and also advanced capabilities for transmission. A burst composed of several packets can be transmitted with precise user-defined inter packet timings. The measuring nodes are synchronized by GPS (Garmin GPS 35 HVS), that provides a PPS (pulse per second) reference signal directly to the DAG card. The accuracy of one-way delay measurements between two ETOMIC nodes is found to be ≈ 0.5 µs, which is mainly limited by the performance of the GPS receivers. This result was obtained in a lab experiment before the deployment of the nodes, where we connected two ETOMIC nodes by an empty link, and transferred a long stream of probe packets between them. Figure 2 shows the histogram of the measured delay, where the bin size reflects the time resolution of the DAG-cards.

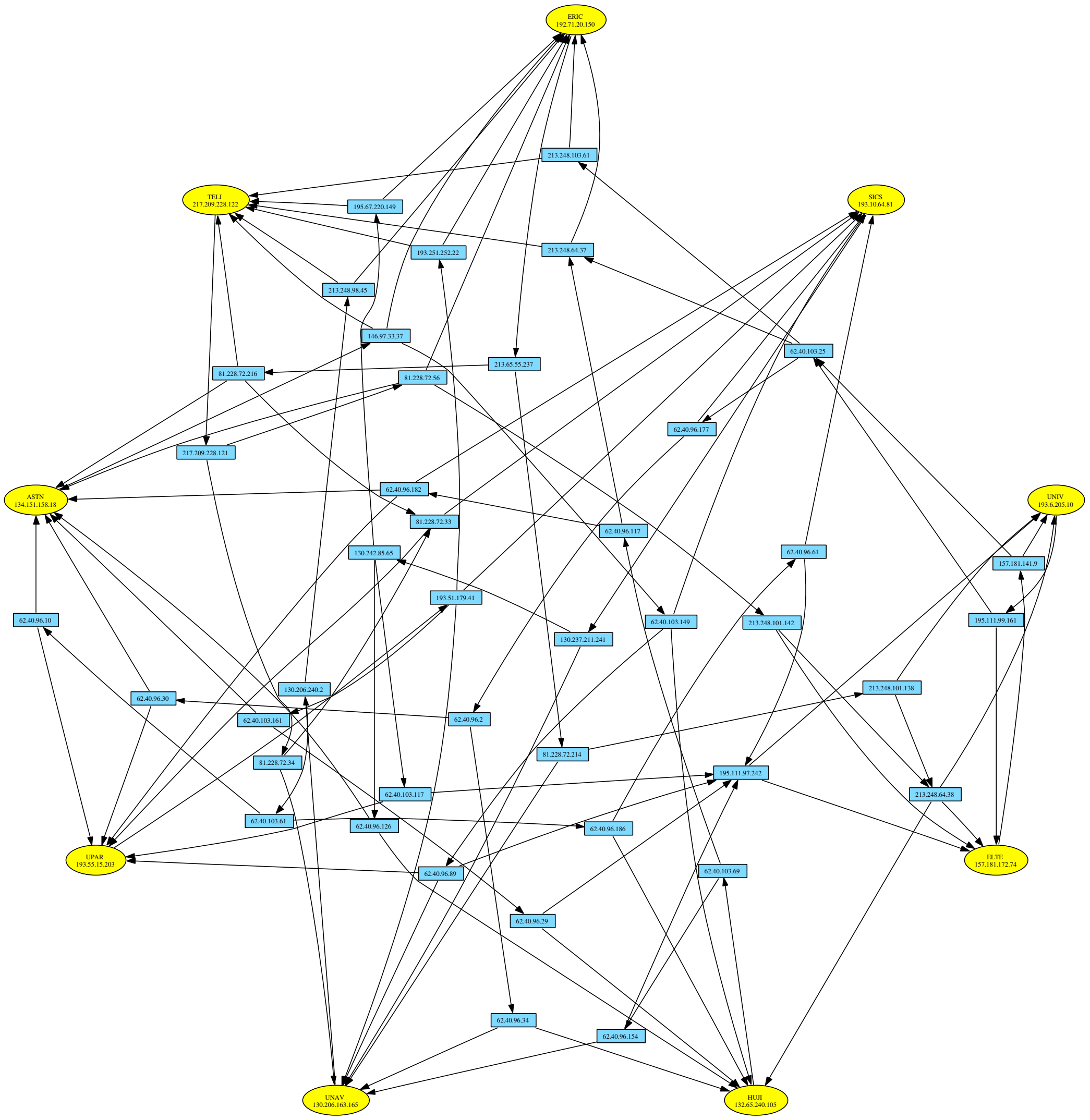

**Fig. 3.** The connection topology between 9 ETOMIC measurement nodes. The ellipse shaped nodes on the edge are ETOMIC measurement nodes with abbreviations given in Table I., while box-shaped nodes in the interior of the graph are branching nodes. The arrows indicate the direction of probe packet flow on a given network segment. The boxes with an IP address of 62.40.X.X are nodes within GÉANT.

Figure 3 depicts the connection topology between 9 measuring nodes, obtained by extensive traceroute measurements, where we extracted from the detailed traces the segments for which characteristics can be resolved by network tomography measurements. This arrangement involves 104 network segments and 42 branching nodes (among these 19 is situated in the GÉANT multi-Gigabit European academic network [28]) with link speeds ranging from 2 to 15 Gb/s.

### 3. Large-scale Queueing Delay Tomography

Here we report a large-scale tomography measurement that involved 9 ETOMIC measurement nodes, and a connectivity graph consisting of 38 branching nodes and 93 network segments. Among these 9 ETOMIC nodes 8 were simultaneously sources of outgoing back-to-back probe pairs, and receivers of the incoming probe packets, while one of the nodes was only used as a receiver. Each of the source nodes sent probe pairs consisting of small sized UDP packets to all the possible pairs of receivers in a round-robin fashion with an inter-pair time of 1 ms, and repeated this process many times. This procedure finally resulted in data sets, each containing two time-series of end-to-end delays with an approximate length of 10000 elements.

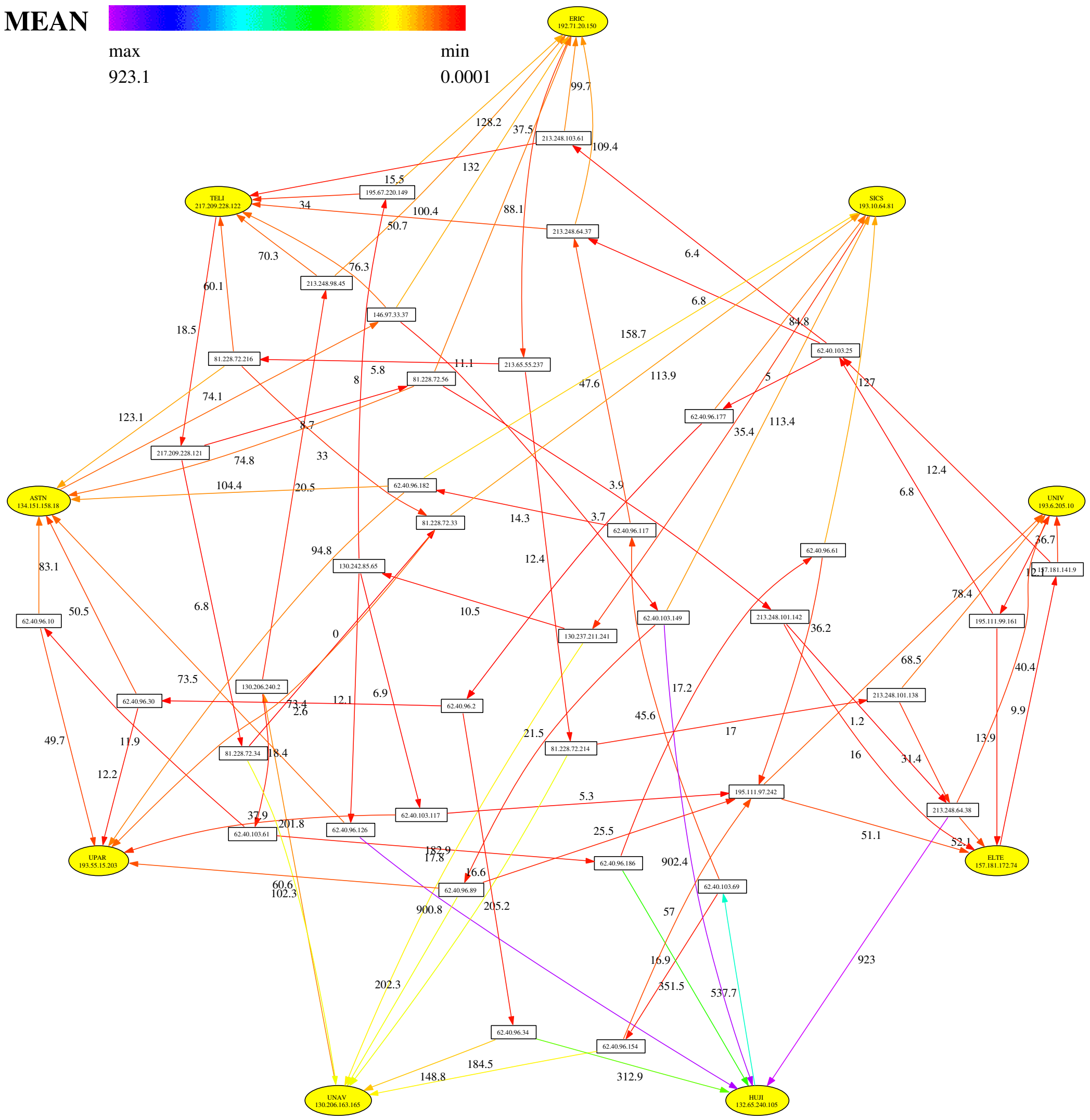

**Fif. 4.** The connectivity graph colored and labeled by the mean queueing delay, given in units of µs, for each network segment.

These data sets comprised the input to the tomography method described in sec. 4 that yielded as an output the queueing delay distributions resolved for each segment contained in the connectivity graph. Since a given segment can be a part of different end-to-end paths, this fact enabled to test the consistency of the results, as well as the averaging of the distributions obtained from different data sets, but attributed for the same segment. For better visualization of the results we extracted the mean and the standard deviation of the queueing delay distributions. These dynamical state parameters are shown on top of the connectivity graph in figs. 4 and 5.

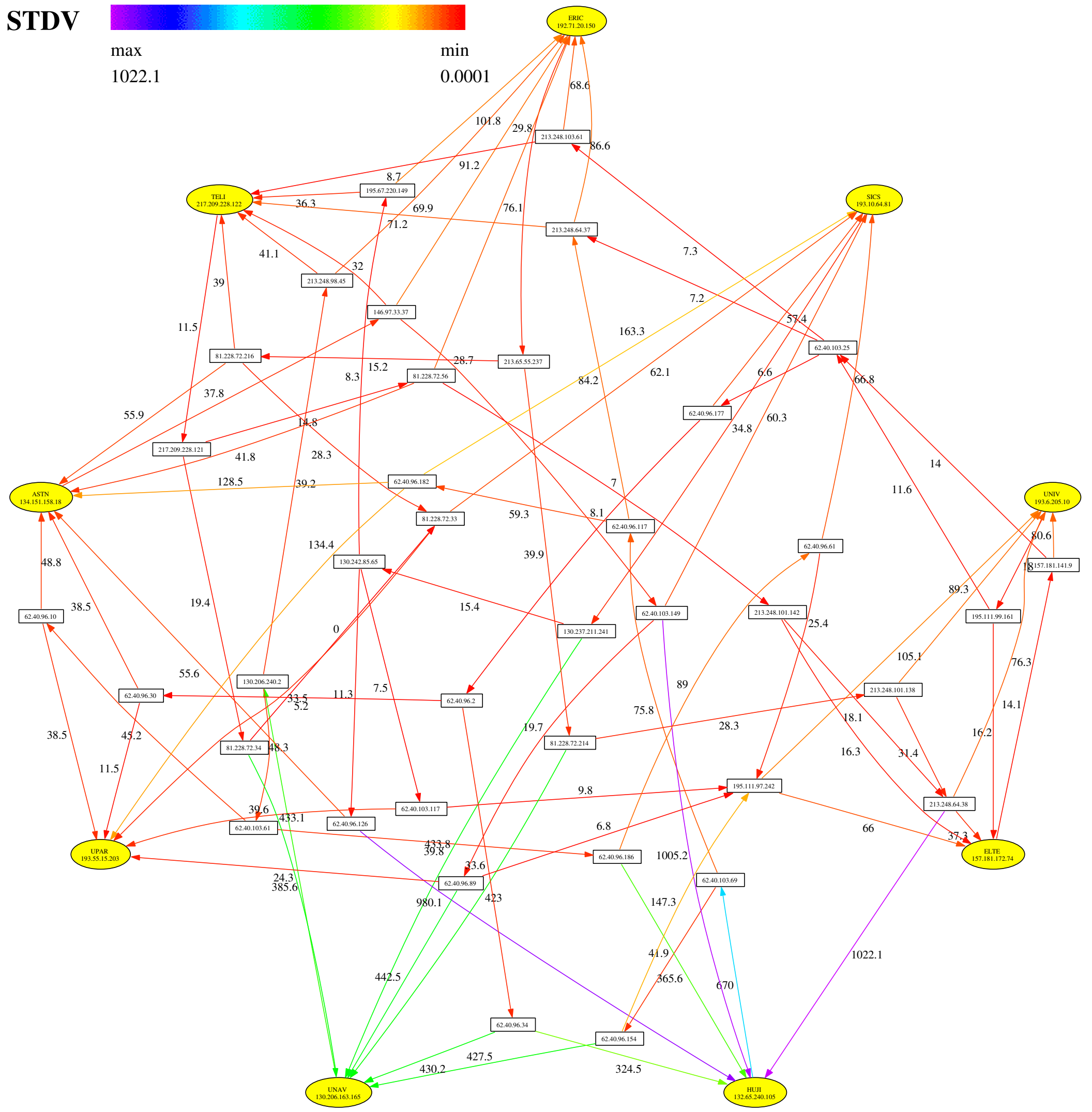

**Fig. 5.** The connectivity graph colored and labeled by the standard deviation of the queueing delays, given in units of µs, for each network segment.

### 3.1. Analysis of the results

The results of figs. 4 and 5 reveal some interesting structure. First of all the state variables of the segments span three orders of magnitude, ranging from the error limit of the delay measurements ( ≈ 0.5 µs), to an average queueing delay of ≈ 1ms, that characterizes a segment which connects GÉANT to the Hebrew University in Jerusalem. The results also reveal an interesting geographical feature, namely that the segments originating or ending in ETOMIC nodes that are located on the south (HUJI, UNAV), are characterized by the highest average and standard deviation of the queueing delays.

As a general feature it can be observed that for all end-nodes incoming segments are characterized by higher values of average queueing delays then outgoing segments. This result can be interpreted by a reasoning that the amount of data downloaded from the Internet to an organization is usually higher, than the amount downloaded from the servers of the organization by clients situated elsewhere in the Internet. Looking at the spatial arrangement of the state variables, one can see that the internal segments that are connections between branching nodes constitute a core which is characterized by the smallest values of the state variables. This is not surprising, since these are network segments in the gigabit backbone.

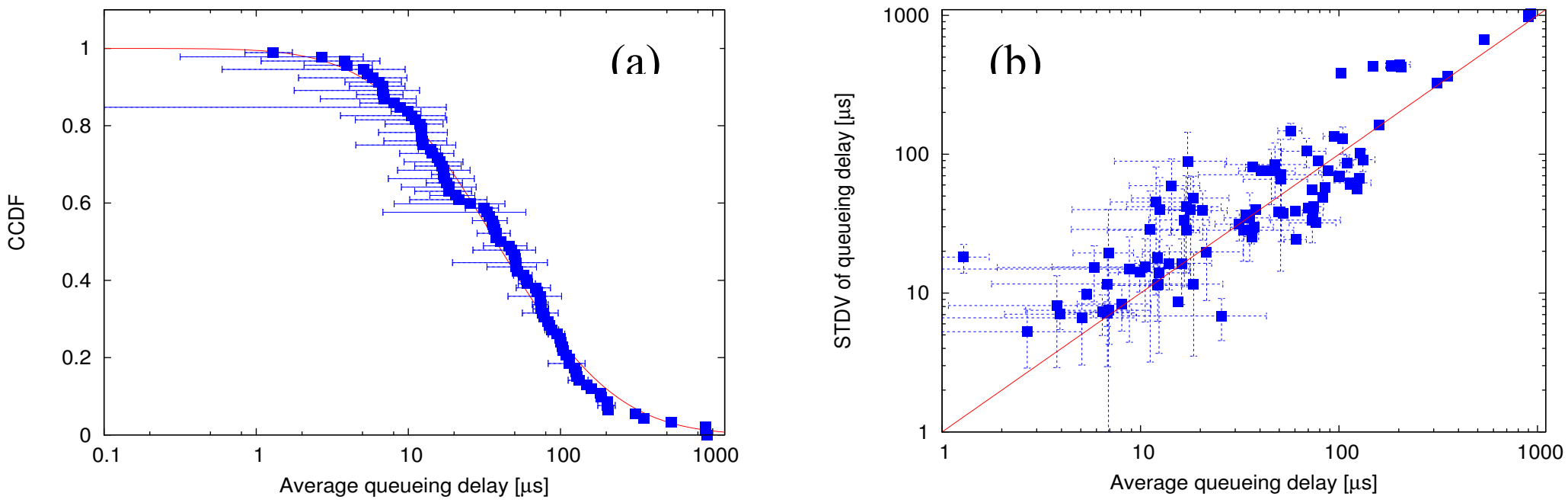

**Fig. 6.** In (a) the complementary cumulative distribution of the average queueing delays are given by filled boxes with error-bars, while the continuous line is a fit given by equation (1). In (b) the standard deviations as a function of the average queueing delays are plotted with error-bars for all the resolved network segments of fig. 4. Here the continuous line marks the diagonal.

To analyze further the results in fig. 6(a) we plot the complementary cumulative distribution function of the average queueing delays on the different segments, that is closely fitted by

$$g(x) = \frac{1}{2}\left[1 - erf\left(\frac{\ln x - m}{\sigma\sqrt{2}}\right)\right], \tag{1}$$

where $x$ is the average queueing delay of a network segment in microseconds, while $\sigma$ and $m$ are the fitting parameters. The implication of the very good fit with $\sigma \approx 1.42$ and $m \approx \ln(37.8\mu s)$ in fig. 6(a) is that the average queueing delay of the different segments follows a log-normal distribution.

$$P(x) = \frac{1}{\sigma\sqrt{2}x} e^{-\frac{(\ln x - m)^2}{2\sigma^2}}. \tag{2}$$

It is interesting that despite the features and spatial structures identified earlier, the data follows a smooth continuous function without segregation of the average queueing delays into clearly visible groups. In fig. 6(b) we plot the standard deviations as a function of the average values of the queueing delay. The figure indicates that the majority of the segments, with a data point near the diagonal, are characterized by a queueing delay distribution that is close to being exponential, while the segments with a large deviation above the diagonal may be characterized by a self-similar traffic-flow and heavy-tailed distributions.

## 4. Description of the Method for Queueing Delay Tomography

In this section we provide the detailed description of the queueing delay tomography method that was applied on the measured end-to-end data in obtaining the results of the previous section.

To achieve a unique labeling of the different network segments of a measurement tree we use the arrangement illustrated in the example of fig. 7. The source node is situated on the top of the hierarchy, labeled with “0”, while branching nodes are put on a given level according to their “distance” from the source. This distance is measured by the number of branching nodes a probe packet must pass to reach the target. Different nodes on the same hierarchical level are enumerated in the incremental order from the left to right. Using this scheme one can identify each node by a unique index, while the segments, indicated by arrows in fig. 7, can be identified by the indices of their target nodes. We denote the one-way delay experienced by the $n$-th probe on the $i$-th segment by $\tilde{X}_i(n)$, and the corresponding end-to-end delay from the source to the $i$-th receiver node by $\tilde{Y}_i(n)$. The queueing delays associated to these quantities are denoted by $X_i(n)$ and $Y_i(n)$. By dropping the $n$ index we indicate the time-series of the respective quantities.

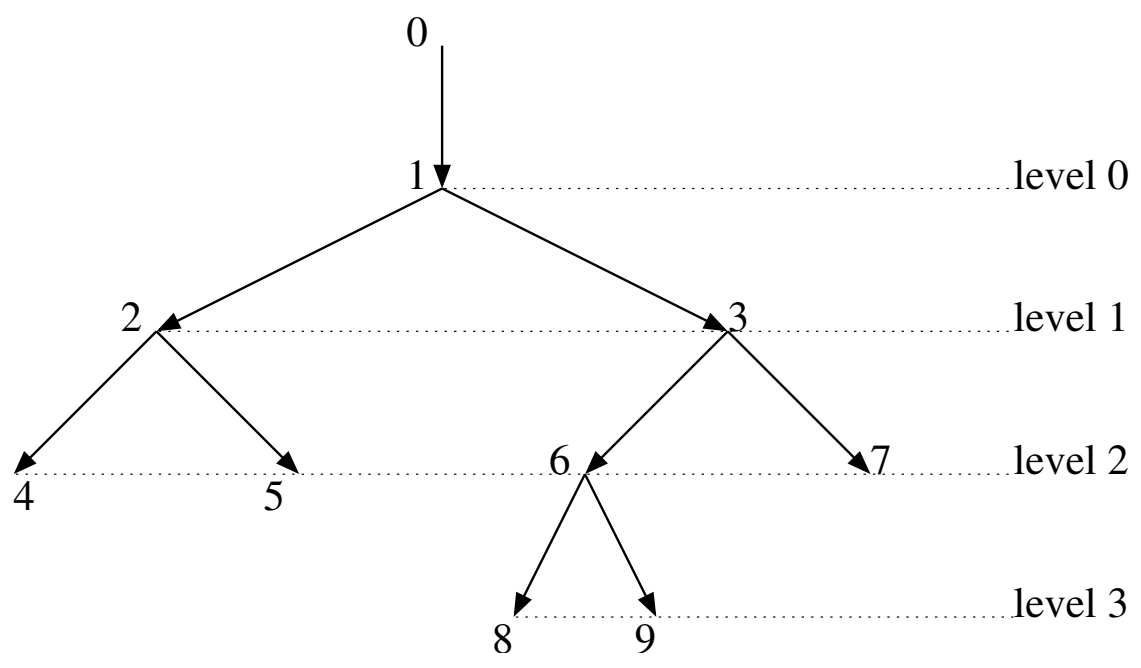

**Fig. 7.** The labeling of the different nodes and network segments of the measurement tree.

### 4. 1. Inference in the two-leaf tree

Here we describe the method of queuing-delay distribution inference in the basic case of the two-leaf tree, which is comprised of a source, two receivers, and a single branching node (see the part of fig. 7 until level 1). The generalization to larger trees will be treated in the next subsection. The known quantities are the time series of end-to-end queueing delays $Y_2$ and $Y_3$ obtained by minimum filtering the $\tilde{Y}_2$ and $\tilde{Y}_3$ time-series

$$Y_2(n) = \tilde{Y}_2(n) - \min(\tilde{Y}_2), \qquad Y_3(n) = \tilde{Y}_3(n) - \min(\tilde{Y}_3). \tag{3}$$

These are measured by a big number of packet pairs, where the packets in a pair are sent back-to-back from the source, and are destined to the two distinct receivers. We count only those pairs where each of the packets has reached its destination, and assign the number of successful pairs by $N$. Above we made the assumption that in every end-to-end delay measurement at least one probe will experience no queueing, thus the minimum can be identified with the propagation delay on the given path, which can be subtracted to yield the delay component due to queueing. Measurement studies of references [29,30], as well as our own experience shows that in the current Internet link utilizations are low enough, so that the above assumption is justified. Using the labeling introduced in the previous subsection we have the trivial relations

$$Y_2(n) = X_1(n) + X_2(n), \qquad Y_3(n) = X_1'(n) + X_3(n), \tag{4}$$

where $X'_1(n)$ stands for the queueing delay on segment 1, experienced by that particular probe in the $n$-th pair, which was destined towards receiver node (3). From this point we will assume perfect correlation of the delays experienced by the two packets in a pair on the common segment $X_1(n) = X'_1(n)$. In practice this ideal situation is approached if for all $n$, $|X'_1(n)-X_1(n)| << X1(n)$. The goal is to estimate the distribution of the unknown $X_1$, $X_2$, and $X_3$ time-series, based on the knowledge of $Y_2$ and $Y_3$.

This can be achieved by introducing the quantized versions of the queueing delays. By using the quantization rule $Y_i^d(n) = jq$, if $(j-1/2)q < Y_i(n) \leq (j+1/2)q$, one can map the end-to-end time-series $Y_i$ to their quantized versions $Y_i^d$, that take values from the set $\{0, q, 2q, \ldots Bq\}$. With the similar quantization we can also introduce the quantized versions of the unknown time-series of $X_i^d$. We choose the quantization parameters $q$ and $B$ so that they satisfy the relation $\mathbf{max}(Y_2, Y_3) < (B+1)q$, which ensures that all the possible values of $Y_i$ and $X_i$ will be included in one of the bins defined by the quantization rule, and thus probabilities can be assigned to the bins. We denote the probability of a queueing delay falling in the $j$-th bin on the $i$-th segment by $P_{i,j}$, and for each $i \in \{1, 2, 3\}$ we have $\sum^{B}_{j=0} P_{i,j}=1$.

The reason behind the introduction of the discrete quantities above was that using the $P_{i,j}$ quantities one can express the probabilities of finding the quantized end-to-end queueing delays in a given bin $P(l,m) \equiv P(Y_2^{d}=lq, Y_3^{d}=mq)$, which can be estimated from the measured data. Assuming temporal and spatial independence of the probe delays these probabilities can be expressed by the convolution

$$P(l,m) = \sum_{k \in H} P_{1,k} P_{2,(l-k)} P_{3,(m-k)}, \tag{5}$$

where the set $H$ is given by $\{B \geq k \geq 0\} \cap \{B \geq (l-k) \geq 0\} \cap \{B \geq (m-k) \geq 0\}$. Labeling the observed variables by $\mathbf{Y} \equiv \{Y_2^{d}(n), Y_3^{d}(n)\}$, $(n \in \{1 \ldots N\})$ and the set of unknown probabilities by $\boldsymbol{\theta} \equiv \{P_{i,j}\}$ finally the unknown probabilities $P_{i,j}$ can be determined from the property that their true value maximizes the log likelihood function

$$\log L(\mathbf{Y} \mid \boldsymbol{\theta}) = \sum_{n=1}^{N} \log P(Y_2^{d}(n), Y_3^{d}(n)), \tag{6}$$

where the probabilities $P(Y_2^{d}(n), Y_3^{d}(n))$ are given by equation (5) evaluated at the known quantized end-to-end queueing delay pairs. Although the log-likelihood-function (6) can not be maximized analytically, there are several numerical procedures to accomplish this goal. Reference [14] suggests to apply the expectation-maximization (EM) algorithm [31].

The EM algorithm is an iterative method to find the maximum likelihood estimate of the unknown probabilities in the presence of some suitably defined hidden or unobserved variables. In our case the hidden variables are the quantized queueing delays on the internal segments of the two-leaf tree, that we denote by $\mathbf{X} \equiv \{X_1^{d}(n), X_2^{d}(n), X_3^{d}(n)\}$, $(n \in \{1 \ldots N\})$. Marginalizing over the hidden variables the condition to maximize equation (6) can be expressed as, for $\forall (i,j)$

$$\frac{\partial \log L(\mathbf{Y} \mid \boldsymbol{\theta})}{\partial P_{i,j}} = \frac{\partial}{\partial P_{i,j}} \left[ \sum_{n=1}^{N} \log \sum_{x \in \mathbf{X}(n)} P(\mathbf{Y}(n), \mathbf{X}(n) \mid \boldsymbol{\theta}) + \sum_{b=1}^{3} \lambda_b \left( 1 - \sum_{j=0}^{B} P_{b,j} \right) \right] = 0, \tag{7}$$

Above the $\lambda_b$ quantities are Lagrange-multipliers to account for the normalization constraint of the $P_{i,j}$ probabilities. Performing the derivations we arrive at

$$\begin{aligned}
&\sum_{n=1}^{N} \frac{1}{\sum_{x \in \mathbf{X}(n)} P(\mathbf{Y}(n), \mathbf{X}(n) \mid \boldsymbol{\theta})} \frac{\partial}{\partial P_{i,j}} \left( \sum_{x \in \mathbf{X}(n)} P(\mathbf{Y}(n), \mathbf{X}(n) \mid \boldsymbol{\theta}) \right) - \lambda_i = \\
&\sum_{n=1}^{N} \left[ \sum_{x' \in \mathbf{X}(n)} \left( \frac{P(\mathbf{Y}(n), \mathbf{X}(n) \mid \boldsymbol{\theta})}{\sum_{x \in \mathbf{X}(n)} P(\mathbf{Y}(n), \mathbf{X}(n) \mid \boldsymbol{\theta})} \frac{\partial}{\partial P_{i,j}} \log P(\mathbf{Y}(n), \mathbf{X}(n) \mid \boldsymbol{\theta}) \right) \right] - \lambda_i = \\
&\sum_{n=1}^{N} \left[ \sum_{x' \in \mathbf{X}(n)} P(\mathbf{X}(n) \mid \mathbf{Y}(n), \boldsymbol{\theta}) \frac{\partial}{\partial P_{i,j}} \log P(\mathbf{Y}(n), \mathbf{X}(n) \mid \boldsymbol{\theta}) \right] - \lambda_i = 0.
\end{aligned} \tag{8}$$

In the above equation the joint probabilities $P(\mathbf{Y}(n),\mathbf{X}(n)\,|\,\boldsymbol{\theta})$ can be factorized as $P(\mathbf{Y}(n)\,|\,\mathbf{X}(n))P(\mathbf{X}(n)\,|\,\boldsymbol{\theta})$. Since $P(\mathbf{Y}(n)\,|\,\mathbf{X}(n))$ does not depend on $\boldsymbol{\theta}$ thus its derivative with respect to $P_{i,j}$ is zero. Using the assumption of spatial and temporal independence we have

$$P(\mathbf{X}(n)\,|\,\boldsymbol{\theta}) = \prod_{i=1}^{3} P(X_i^d(n)), \tag{9}$$

where $P(X_i^d(n))$ is the probability that the $n$-th probe experienced a quantized queueing delay of $X^d$ on the $i$-th segment. Substituting equation (9) into (8) we may write

$$\begin{aligned}&\sum_{n=1}^{N}\left[\sum_{x'\in\mathbf{X}(n)} P(\mathbf{X}(n)\,|\,\mathbf{Y}(n),\boldsymbol{\theta})\frac{\delta(X_i^d(n)=jq)}{P_{i,j}}\right]-\lambda_i = \\ &\frac{1}{P_{i,j}}\sum_{n=1}^{N} P(X_i^d(n)=jq\,|\,\mathbf{Y}(n),\boldsymbol{\theta})-\lambda_i = 0,\end{aligned} \tag{10}$$

After rearranging and using the normalization constraints $\sum_{j=0}^{B} P_{i,j}=1$, finally we arrive at the following system of coupled equations. For $\forall(i,j)$

$$P_{i,j} = \frac{1}{N}\sum_{n=1}^{N} P(X_i^d(n)=jq\,|\,\mathbf{Y}(n),\boldsymbol{\theta}), \tag{11}$$

where $P(X_i^d(n)=jq\,|\,\mathbf{Y}(n),\boldsymbol{\theta})$ stands for the conditional probability of $X_i^d(n)=jq$, given the observed values $Y_2^d(n), Y_3^d(n)$ of the n-th packet pair. According to Bayes-law these conditional probabilities can be expressed as

$$\begin{aligned}P(X_1^d(n)=jq\,|\,Y_2^d(n)=lq, Y_3^d(n)=mq,\boldsymbol{\theta}) &= \frac{P_{1,j}P_{2,(l-j)}P_{3,(m-j)}}{P(l,m)}, \\ P(X_2^d(n)=jq\,|\,Y_2^d(n)=lq, Y_3^d(n)=mq,\boldsymbol{\theta}) &= \frac{P_{1,(l-j)}P_{2,j}P_{3,(m-l+j)}}{P(l,m)}, \\ P(X_3^d(n)=jq\,|\,Y_2^d(n)=lq, Y_3^d(n)=mq,\boldsymbol{\theta}) &= \frac{P_{1,(m-j)}P_{2,(l-m+j)}P_{3,j}}{P(l,m)},\end{aligned} \tag{12}$$

where $P(l,m)$ is given by equation (5). The system of coupled equations (5,11,12) can be solved iteratively. One starts by assigning initial values for all the $P_{i,j}$ probabilities (e.g. the evenly distributed probabilities $P_{i,j}^0 = 1/(B+1)$), then equations (11) together with (5) and (12) provide the improved estimates $P'_{i,j}$. This process is iterated until a suitable criterion for convergence is met. We terminate the iterations if for $\forall(i,j)$

$$\left|P'_{i,j}-P_{i,j}\right|B \le \varepsilon \tag{13}$$

where $\varepsilon$ is a chosen small number, e.g. 0.0001.

### 4. 2. Inference in arbitrary trees

For an arbitrary large tree it can be shown that the queueing delay distribution on any segment can be either estimated directly by the inference algorithm developed for the two-leaf tree, or can be generated by numerical deconvolution of two directly inferred distributions. To resolve all the network segments by a minimal approach one must perform packet-pair measurements in a way that all the branching nodes and all the receivers are visited at least in one of the measurements. We demonstrate this in the following example.

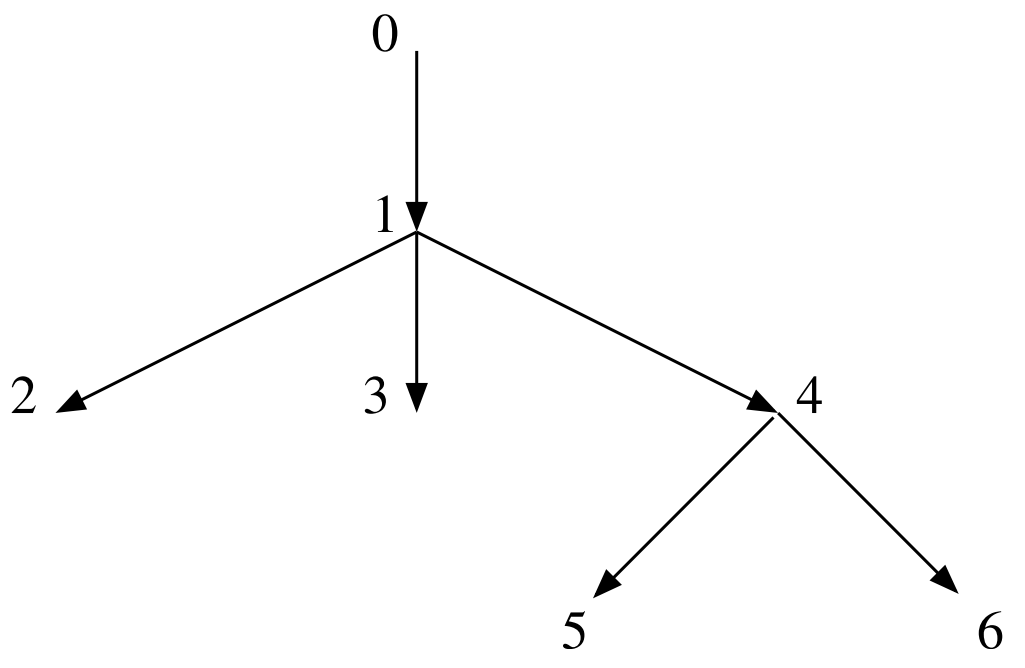

**Fig. 8.** An example of a measurement tree with 4 receiver nodes.

Consider the tree shown in fig. 8. Performing the back-to-back packet pair measurements to the receiver pairs of (2,3) and (5,6) in a round robin fashion, and applying the inference algorithm developed for the two-leaf tree estimates directly the probabilities $P_{1,j}$, $P_{2,j}$, $P_{3,j}$ and of $P_{(0\to4),j}$, $P_{5,j}$, $P_{6,j}$ ($j \in \{0,1,\ldots B\}$). Here we denoted by (0→4) the common segment originating in 0 and ending in branching node 4. The probabilities $P_{(0\to4),j}$ can be expressed by the convolution

$$P_{(0\to4),j} = \sum_{k\in G} P_{1,(j-k)} P_{4,k}, \tag{14}$$

Where the set $G$ is given by $\{B\geq k\geq 0\} \cap \{B\geq (j-k)\geq 0\}$. This convolution can also be viewed as a matrix operation

$$P_{(0\to4),j} = \sum_{k=0}^{B} A_{j,k} P_{4,k}, \tag{15}$$

where the matrix element is $A_{j,k} = P_{1,(j-k)}$, if $k \in G$, and $A_{j,k} = 0$, if $k \notin G$,. This way the problem of numerical deconvolution to obtain the remaining probabilities on segment 4 can be mapped to a matrix inversion problem. For this task the best suited numerical method is the non-negative least squares (NNLS) algorithm [32], which guarantees that the resulting probabilities will be strictly non-negative.

In reference [33] we have investigated the performance of the inference method described so far in extensive realistic simulation studies, as well as in controlled LAN experiments with ETOMIC measurement nodes. These investigations revealed impressive agreement of the estimated and the real queueing delay distributions, and an error in the estimation of the first moments being less then the value of the chosen bin size.

### 5. Conclusion

This paper presented dynamical state measurements of a part of the European Internet, conducted via the ETOMIC measurement infrastructure, using special active probing techniques and the methods of network tomography. This very precise and fully synchronized infrastructure meets the requirements needed to perform large-scale unicast tomography measurements, and can be viewed as the prototype of network testbeds, that will be able to operate in the uncooperative Internet of the future. In the paper we have investigated a particularly important state variable of the Internet, the characteristics of queueing delay distributions on different network segments, which provides information about traffic properties and the state of congestion in the network. As the main results we presented maps of the averages and standard deviations of queueing delay, and identified in them various structures. We find that the average queueing delay of network segments spans three orders of magnitude, and its distribution function closely follows a log-normal distribution.

**Acknowledgements**
The authors thank the partial support of the National Office for Research and Technology (NKFP 02 032 04) and the European Union IST FET COMPLEXITY EVERGROW Integrated Project.